\documentclass[aps,twocolumn,pra,floatfix,longbibliography,superscriptaddress,amssymb]{revtex4-2}
\pdfoutput=1

\usepackage{amsmath,bbm}
\usepackage{graphicx}
\usepackage{multirow}
\usepackage{hyperref}
\usepackage{amsfonts}
\usepackage{amsbsy}
\usepackage{xcolor}
\usepackage{soul}
\usepackage{graphicx}
\usepackage{mathtools}
\usepackage{bm}
\usepackage[normalem]{ulem}

\renewcommand{\vec}[1]{\mathbf{#1}}

\newcommand{\Pc}{\mathcal{P}}
\newcommand{\cbE}{\boldsymbol{\mathbf{\cal E}}}

\newcommand{\unitvec}[1]{\hat{\mathbf{{#1}}}}
\renewcommand{\vec}[1]{\mathbf{#1}}

\newcommand{\vecg}[1]{\bm{#1}}

\definecolor{mygreen}{RGB}{20,148,20}

\usepackage{filemod}

\definecolor{mygreen}{RGB}{34 139 34}

\begin{document}

\title{Optical magnetism and wavefront control by arrays of strontium atoms }

\author{K.~E.~Ballantine}
\affiliation{Department of Physics, Lancaster University, Lancaster, LA1 4YB, United Kingdom}
\author{D. Wilkowski}
\affiliation{Nanyang Quantum Hub, School of Physical and Mathematical Sciences, Nanyang Technological University, 21 Nanyang Link, Singapore 637371, Singapore}
\affiliation{MajuLab, International Joint Research Unit IRL 3654, CNRS, Universit\'e C\^ote d'Azur, Sorbonne Universit\'e, National University of Singapore, Nanyang
Technological University, Singapore}
\affiliation{Centre for Quantum Technologies, National University of Singapore, 117543 Singapore, Singapore}
\author{J.~Ruostekoski}
\affiliation{Department of Physics, Lancaster University, Lancaster, LA1 4YB, United Kingdom}

\date{\today}

\begin{abstract}
   By analyzing the parameters of electronic transitions, we show how bosonic Sr atoms in planar optical lattices can be engineered to exhibit optical magnetism and other higher-order electromagnetic multipoles that can be harnessed for wavefront control of incident light. Resonant $\lambda\simeq 2.6\mu$m light for the $^3D_1\rightarrow {^3}P_0$ transition mediates cooperative interactions between the atoms while the atoms are trapped in a deeply subwavelength optical lattice. The atoms then exhibit collective excitation eigenmodes, e.g., with a strong cooperative magnetic response at optical frequencies, despite individual atoms having negligible coupling to the magnetic component of light. We provide a detailed scheme to utilize excitations of such cooperative modes consisting of arrays of electromagnetic multipoles to form an atomic Huygens' surface, with complete $2\pi$ phase control of transmitted light and almost no reflection, allowing nearly arbitrary wavefront shaping. In the numerical examples, this is achieved by controlling the atomic level shifts of Sr with off-resonant ${^3P}_J\rightarrow {^3D}_1$ transitions, which results in a simultaneous excitation of arrays of electric dipoles and electric quadrupoles or magnetic dipoles. We demonstrate the wavefront engineering for a Sr array by realizing the steering of an incident beam and generation of a baby-Skyrmion texture in the transmitted light via a topologically nontrivial transition of a Gaussian beam to a Poincar\'{e} beam, which contains all possible polarizations in a single cross-section.
    
\end{abstract}

\maketitle

\section{Introduction}

Extreme wavefront control is the ultimate goal of manipulating light by material interfaces. However, the quest for precise optical control is often limited by vanishingly weak couplings of natural polarizable media to the magnetic component of light~\cite{LandauED}. This shortcoming has driven the development of metamaterials and metasurfaces~\cite{Zheludev12,Yu14,Luo18}. Thin artificial metasurfaces, in particular, can impart abrupt phase changes on transmitted and reflected light, leading to the development of Huygens' surfaces~\cite{Pfeiffer13,Decker15,Yu15,Chong15,Shalaev15}. These surfaces are physical realizations of Huygens' principle, which states that each point in a propagating beam acts as an independent source of forward-propagating waves~\cite{Huygens}, with a more rigorous formulation modeling these sources as crossed electric and magnetic dipoles~\cite{Love1901}. The resulting metasurfaces allow the phase, polarization, and direction of transmitted light to be engineered. However, metasurface functionalities have limitations: Fabrication inconsistencies lead to inhomogeneous resonance broadening, and absorptive losses are hard to avoid at optical frequencies. In addition, they almost entirely operate in the classical regime. 

Meanwhile, atoms trapped in a periodic subwavelength planar optical lattice have been shown in experiments to exhibit collective, spatially delocalized subradiant optical excitations~\cite{Rui2020}, and related experiments on collective excitations have also been performed in other periodic structures~\cite{Jenkins17,Ferioli21}. Cooperatively responding optical systems of subwavelength atomic arrays have inspired a large body of theoretical studies, with examples including manipulation of subradiance~\cite{Facchinetti16,Asenjo-Garcia2017a,Jen17,Guimond2019},  single-photon storage~\cite{Ballantine21quantum,Rubies22,Ballantine21bilayer}, atom and excitation statistics~\cite{Zhang2018,Orioli19}, optical cavity-like phenomena~\cite{Bettles20,Parmee20bistable,Parmee2020}, collective antibunching~\cite{Cidrim20,Williamson2020b,Holzinger21}, connected arrays~\cite{Yoo20}, optomechanics~\cite{Shahmoon19}, and parity-time symmetry breaking~\cite{Ballantine21PT}. 
In particular, it was recently shown how a bilayer array of atoms could form a Huygens' surface via emerging collective excitations that mimic an array of crossed electric and magnetic dipoles, even when the atoms only undergo electric dipole transitions~\cite{Ballantine20Huygens,Ballantine21wavefront}.

Although the optical transmission experiment~\cite{Rui2020} was performed with alkali-metal atoms trapped in a Mott insulator state of an optical lattice, there are several other promising atomic species. In particular, alkaline-earth-metal and rare-earth-metal atoms provide great flexibility in the experimental control of their optical properties~\cite{Daley08,Fukuhara09,Ye08}, and Mott insulator states of $^{174}$Yb and $^{84}$Sr have been experimentally realized~\cite{Fukuhara09, Stellmer12}.
Sr and Yb exhibit many optical transitions, and their narrow optical resonances find applications, e.g., in atomic clocks~\cite{Bothwell19} and in fermionic many-body physics~\cite{Fukuhara07}. In the bosonic isotopes of Sr and Yb, the nuclear spin vanishes, and the atoms provide prototype models for spatially isotropic $J=0\rightarrow J^\prime=1$ optical transitions. Moreover, Sr exhibits a long-wavelength optical transition together with different magic wavelength resonances that provide optical lattice spacing much smaller than the probe resonance wavelength~\cite{Olmos13}.

Here we demonstrate how a long-standing challenge of generating optically active magnetism in atomic media can be implemented by analyzing in detail the specific electronic properties of bosonic Sr atoms. We show how cooperative responses of atoms exhibiting arrays of synthetic magnetic dipoles and electric quadrupoles are formed from electric dipole transitions and how these can be utilized for constructing a Huygens' surface of atoms. We analyze the $\lambda=2.6\mu$m optical transition from the $5s4d {}^3D_1$ to the metastable $5s5p {}^3P_0$ state. The coupling of light to other than electric dipole transitions of an individual Sr atom is negligible. However, due to cooperative light-mediated interactions, collective excitation eigenmodes emerge in a bilayer array where the atomic dipoles can point in different directions. For one such mode, radiative transitions form an approximate circle of electric dipoles around a four-atom unit cell in each of the unit cells of the array. The cooperatively coupled electric dipole transitions of atoms at the corners of the unit cell then mimic an approximate loop of oscillating current, producing a magnetic dipole. 
With a different orientation of the electric dipole transitions each unit cell exhibits an electric quadrupole.
The scattering symmetries of these higher-order multipoles in the forward and backward directions are different from those of electric dipoles~\cite{Liu17mm}. In a bilayer array, they can then interfere with electric dipoles in a planar array by providing complete transmission with no reflection to realize a Huygens' surface.

By applying carefully chosen control fields to the Sr atoms, we use in a numerical example off-resonant coupling to higher $5s6p {}^3P_J$ states to shift the $^3D_1$ atomic level energies and control the simultaneous excitation of electric-dipole and electric-quadrupole collective modes. This generates a Huygens' surface with $\gtrsim 89\%$ transmission and complete phase control of the light field for a $31\times31\times4$ lattice. We analyze the sensitivity of the wavefront shaping to experimental errors and show that an offset in one parameter can be compensated for by varying the remaining values; e.g., when the phase of the off-resonant control field is allowed to vary, an efficient Huygens' surface with high transmission ($>80\%)$ can be maintained by also varying the control-laser intensity. 

Using the Huygens' surface, we demonstrate beam steering, changing the direction of the transmitted beam relative to the incident beam, which allows for optical sorting of light for subsequent processing. Moreover, we show how the Sr array can transform a topologically trivial Gaussian beam into a Poincar\'{e} beam, containing all possible polarizations in a single cross-section~\cite{Beckley10}. The Stokes vector of this beam provides a non-trivial topological mapping from the transverse plane to the Poincar\'{e} sphere, forming a baby-Skyrmion texture~\cite{Donati2016,Gao2020}.

\section{Sr atom and optical lattices}

\begin{figure}[t]
  \centering
   \includegraphics[width=0.7\columnwidth]{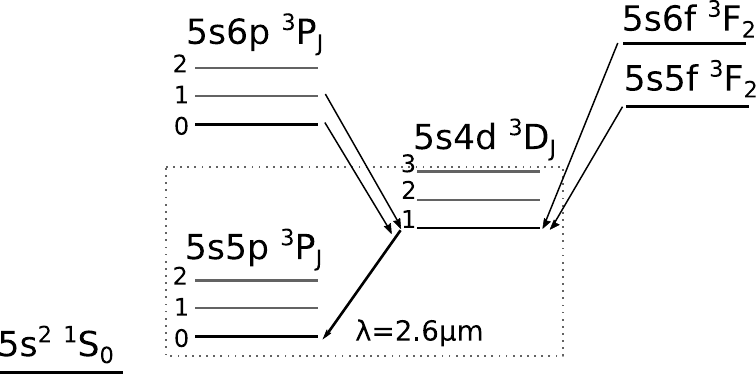}
  \caption{Selected relevant levels and transitions of $^{88}$Sr. While the overall ground state is the singlet $^1S_0$ state, low-lying triplet states form a separate series, with a $^3D_1\rightarrow\,^3P_0$ transition with wavelength $2.6\mu{\rm m}$. Further transitions are used to trap the atoms at a magic wavelength, and to induce controllable ac Stark shifts. }
  \label{fig:level_scheme}
\end{figure}

We show selected levels and transitions for a Sr atom in Fig.~\ref{fig:level_scheme}. The ground state is the singlet state $5s^2\,^1S_0$. However, transitions between singlet and triplet states are electric dipole forbidden in the absence of magnetic fields. The lowest triplet state $5s5p\,^3P_0$, therefore, forms a metastable effective ground state, leading to a $J=0\rightarrow J^\prime=1$ transition with the $5s4d\,^3D_1$ states, with a wavelength $\lambda\simeq 2.6\mu{\rm m}$ and $\gamma\simeq 1.45\times 10^{5} {\rm s}^{-1}$~\cite{Zhou10,Werij92} where $\gamma=\mathcal{D}^2 k^3/(6\pi\epsilon_0\hbar)$ is Wigner-Weisskopf linewidth with $\mathcal{D}$ the reduced matrix dipole element and $k=2\pi/\lambda$. Additional resonances~\cite{Olmos13} are used to trap the atoms in an optical lattice at a magic wavelength. Trapping at the magic wavelength has the advantage of being robust against fluctuations of intensity and atomic positions~\cite{Katori03}. However, atoms could also be strongly trapped in deep optical lattices in the Lamb-Dicke regime at other wavelengths.

\subsection{Level polarizabilites}

To find the condition for magic wavelength trapping in Sec.~\ref{sec:trapping}, and especially to control the atomic level shifts of the $m=0,\pm 1$ sublevels of the $^3D_1$ state to engineer the Huygens' surface in Sec.~\ref{sec:HSshifts}, we require the general expression of the light shift of the atomic energy of level $i$, for off-resonant coupling to level $k$. In the presence of a plane wave having an intensity $I$, frequency $\omega$, and polarization $\unitvec{e}_q$ for $q=0,\pm1$, this is given by~\cite{Grimm95,Boyd1B} 
\begin{equation} \label{eq:LF}
    \delta\omega_i=-\frac{6 c^2}{\hbar\omega^3}\frac{I}{2J_i+1}\sum_k \gamma_{ik}\omega_{ik}(2J_k+1)\frac{\langle J_im_i\vert J_k1m_kq\rangle^2}{w_{ik}^2-w^2},
\end{equation} 
where $J_i$ ($J_k$) is the total angular momentum quantum number of the ground (excited) state, $\langle J_im_i\vert J_k1m_kq\rangle$ the corresponding (real) Clebsch-Gordan coefficient, $\omega_{ik}$ the frequency and $\gamma_{ik}$ the linewidth of the transition. The light shift for arbitrary direction and polarization can be obtained from a geometric decomposition of this shift into scalar, vector, and tensor polarizabilities (see Appendix~\ref{sec:shifts}).

\subsection{Magic wavelength and trapping geometry}
\label{sec:trapping}

\begin{figure}[t]
  \centering
   \includegraphics[width=0.7\columnwidth]{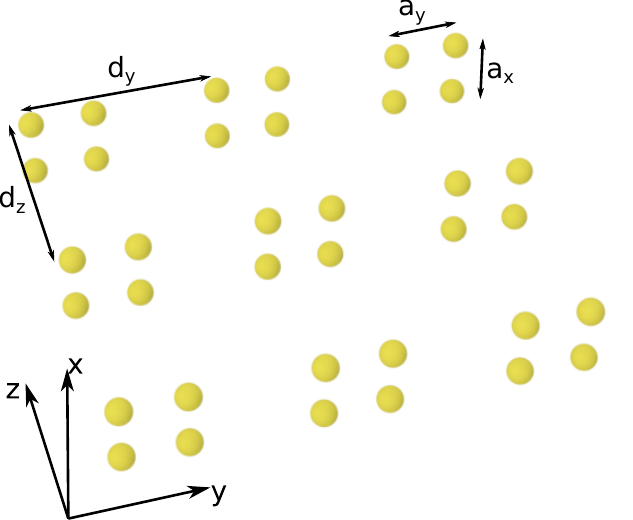}
  \caption{Geometry of an atomic bilayer consisting of square unit cells. The bilayer forms a regular rectangular lattice in the $yz$ plane with lattice constants $d_y$, $d_z$. At each lattice site four atoms form a square unit cell with side-length $a_x$ and $a_y$ in the $x$ and $y$ direction, respectively. }
  \label{fig:sq1}
\end{figure}

\begin{figure}[ht]
\includegraphics[width=0.5\textwidth]{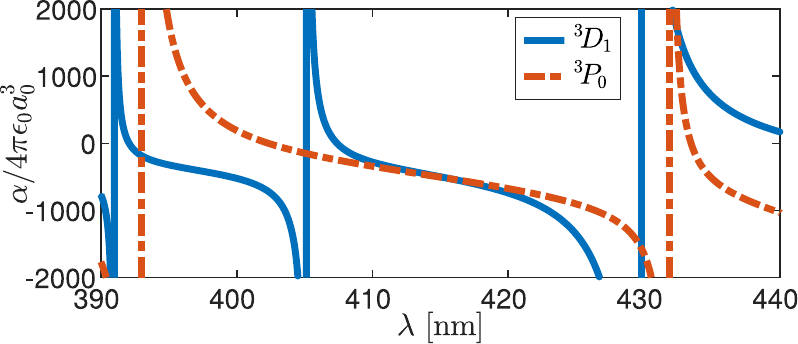}
\caption{Wavelength dependence of the polarizability of the $5s4d\,^3D_1$ and $5s5p\,^3P_0$ states around $415\,$nm, close to resonance with the $5s5f\,^3F_2$ and $5s6f\,^2F_2\rightarrow 5s4d\,^3D_1$ transitions. We note a magic wavelength with negative polarizability at $415\,$nm.} \label{400} 
\end{figure}

The atoms are trapped in a rectangular lattice in the $yz$ plane with lattice constants $d_y$ ($d_z$) in the $y$ ($z$) directions. Each site consists of a four-atom unit cell, with each atom at a fixed position forming a square oriented in the $xy$ plane with side-lengths $a_x$ ($a_y$) in the $x$ ($y$) directions. 
To find realistic and feasible values for subwavelength trapping, we first explore the magic wavelength condition for the transition $5s5p\,^3P_0\rightarrow 5s4d\,^3D_1$, which provides a very short wavelength lattice solution for Sr~\cite{Olmos13}. The polarizabilities of the two states as a function of wavelength are given in Fig.~\ref{400}. A particularly suitable magic wavelength solution was calculated in Ref.~\cite{Olmos13} at $\lambda_m\sim 415\,$nm. While the polarizability of the $^3P_0$ level is well-known because of its metrological interest~\cite{Zhou10}, the transitions from $^3D_1$, and especially their decay rates, are much less frequently explored~\cite{Sansonetti10}. Hence, the appearance of the precise value of this magic wavelength has to be taken with great care.

For the polarizability computation of the $^3D_1$ level, we identified two important transitions given in Table~\ref{table2}. The magic wavelength $\lambda_m$ allows for homogeneous trapping of atoms at the intensity minima of a standing wave with a period $a\geq \lambda_m/2\approx 207$nm, allowing for deeply subwavelength spacing $a/\lambda\simeq 0.08$ when compared to $\lambda\simeq 2.6\,\mu$m, the resonance wavelength of the $^3D_1\rightarrow$ $^3P_0$ transition. In comparison, recent experiments with $^{87}$Rb achieved a ratio of spacing $a/\lambda\simeq 0.68$~\cite{Rui2020}. While this was sufficient to demonstrate the cooperative response, with subradiant collective decay rates below the fundamental limit of a single isolated atom, Sr arrays allow for multiple atoms within a subwavelength spacing.

\begin{table}[ht]
\caption{Wavelengths and linewidths of selected transitions used for magic-wavelength trapping of the $^3D_1$ and $^3P_0$ states.} 
\centering 
\begin{tabular}{c c c} 
\hline\hline 
Transition & Wavelength ($nm$) & $\gamma$ ($s^{-1}$) \\ [0.5ex] 
\hline 
$5s5f\,^3F_2\rightarrow 5s4d\,^3D_1$ & $430.9$ & $2\times 10^{7}$ \\ 
$5s6f\,^3F_2\rightarrow 5s4d\,^3D_1$ & $406.2$ & $5\times 10^{6}$ \\  [0.5ex] 
\hline 
\end{tabular}
\label{table2}
\end{table}

A 3D optical trap with independent control over the spacing in each direction can be formed by varying the angle of the trapping beams, generating an accordion lattice~\cite{Morsch01,Assam10,Hutson19}. 
Here, we use three pairs of laser beams, with each pair interfering to create a standing wave in the $x$, $y$, and $z$ directions. Each laser is tuned approximately to the magic wavelength, such that all atoms experience an identical trapping potential regardless of internal state, with a slight offset in detuning to average out unwanted interference effects (see Appendix \ref{sec:trap}).   
The result is a 3D lattice with spacing $a_x,a_y,d_z \agt 0.08\lambda$. The bilayer geometry with square unit cells can then be implemented by selectively filling sites of this lattice, with $d_y=n a_y$ for integer $n$. 
While the atoms are trapped here at the magic wavelength, this is not a strict requirement as, for sufficient trapping strength, the atoms will not be inhomogeneously broadened in the Lamb-Dicke regime. 
In this case, the geometry could alternatively be realized by a 2D regular square lattice combined with a double-well super-lattice in the $x$ direction~\cite{Kangara18,Koepsell20,Koutentakis20,Gall21}, or with optical tweezers~\cite{Barredo16}, including at blue detuning using an optical bottle trap~\cite{Xu10}.

\section{Collective light-atom interactions}
\label{sec:LLI}

When the intensity of the incident field is low, the response is linear, and the atoms behave as a set of coupled, driven harmonic oscillators~\cite{Lee16,Morice1995a,Ruostekoski1997a,Sokolov2011}. 
Furthermore, when the interatomic separation is sufficiently small, light is scattered between atoms multiple times before escaping. Each atom is driven not just by the external field but also by that from every other atom, leading to strong dipole-dipole interactions and cooperative responses.
We now describe the theory for calculating the collective optical response of an ensemble of $N$ atoms at positions $\vec{r}_i,\ldots,\vec{r}_N$ and with a $J=0\rightarrow J^\prime=1$ transition to a coherent incident field $\cbE(\vec{r})$ in the limit of low light intensity. 
The evolving system is described by the polarization amplitudes $\Pc_{\sigma}^{(j)}$ with dipole moment 
\begin{equation}\label{eq:dipole}
\vec{d}_j = \mathcal{D}\sum_{\sigma} \Pc_{\sigma}^{(j)}\unitvec{e}_\sigma,
\end{equation}
where $\mathcal{D}$ denotes the reduced dipole matrix element and $\unitvec{e}_\sigma$ is the associated unit vector.
Atoms occupy only the ground state, with changes in $\Pc_{\sigma}^{(j)}$ linearly proportional to the incident field. 
The low light intensity limit is obtained by deriving the equations of motion to first order in $\cbE(\vec{r})$ by keeping terms proportional to at most one of $\Pc_{\sigma}^{(j)}$ and $\cbE(\vec{r})$, and also neglecting terms proportional to the excited state populations.
The polarization amplitudes obey
\begin{equation}\label{eq:Peoms}
\frac{d}{dt} \Pc_{\sigma}^{(j)}  
  =  \left( i \Delta_\sigma^{(j)} - \gamma \right)
  \Pc^{(j)}_\sigma + i\frac{\xi}{\mathcal{D}}\hat{\vec{e}}_\sigma^{\ast}\cdot\epsilon_0\vec{E}_{\rm ext}(\vec{r}_j),
\end{equation}
with $\Delta_{\sigma}^{(j)}$ the shift of level $\sigma$ on atom $j$, $\gamma=\mathcal{D}^2k^3/(6\pi\epsilon_0\hbar)$ the single-atom linewidth, and $\xi=6\pi\gamma/k^3$.
The external field $\vec{E}_{\rm ext}$ experienced by each atom, 
\begin{equation}\label{eq:Eext}
\vec{E}_{\rm ext}(\vec{r}_j) = \cbE(\vec{r}_j) + \sum_{l\neq j} \vec{E}_s^{(l)}(\vec{r}_j),
\end{equation} 
is the sum of the incident field and the scattered field from each other atom $l$, $\vec{E}_s^{(l)}(\vec{r})$, given by
\begin{equation}
    \epsilon_0 \vec{E}_s^{(l)}(\vec{r})=\mathsf{G}(\vec{r}-\vec{r}_l)\vec{d}_l
\end{equation}
where $\mathsf{G}(\vec{r})$ is the dipole radiation kernel~\cite{Jackson}. 
Inserting Eq.~\eqref{eq:Eext} into Eq.~\eqref{eq:Peoms}, the equations of motion can be rearranged in vector form~\cite{Lee16} as
\begin{equation}
\dot{\vec{b}}  = i\mathcal{H}\vec{b}+\vec{f},
\end{equation}
where $\vec{b}$ and $\vec{f}$ are vectors of the polarization amplitudes $\Pc_{\sigma}^{(j)}$ and drive $i(\xi/\mathcal{D})\unitvec{e}_\sigma^\ast\cdot\epsilon_0\cbE(\vec{r}_j)$, respectively. The matrix $\mathcal{H}$ describes interactions between different atoms due to multiple scattering of light, with off-diagonal elements given by $ \xi \unitvec{e}_\sigma^\ast\cdot\mathsf{G}(\vec{r}_j-\vec{r}_k)\unitvec{e}_{\sigma^\prime}$, leading to cooperative responses~\cite{guerin16_review,JavanainenMFT,Andreoli21}. The diagonal elements are $\Delta_{\sigma}^{(j)}+i\gamma$, where $\Delta_\sigma^{(j)}=\delta_\sigma^{(j)}+\Delta$ consists of an overall detuning $\Delta=\omega-\omega_0$ of the laser frequency $\omega$ from the single-atom resonance $\omega_0$, plus a relative shift $\delta_\sigma^{(j)}$ of each level. The response of the array can be understood through the eigenvectors $\vec{v}_n$ and eigenvalues $\delta_n+i\upsilon_n$ of $\mathcal{H}$~\cite{Sokolov2011} giving the collective line shifts $\delta_n$ and linewidths $\upsilon_n$~\cite{Jenkins_long16}.

\section{Engineering optical magnetism and higher multipole moments with strontium}
\label{sec:magnetism}

Optical magnetism has been shown to emerge from the collective response of closely spaced atoms, as two~\cite{Alaee20} or four~\cite{Ballantine20Huygens} atom unit cells act as magnetic antennae. We now show how the optical transitions of bosonic Sr can be used to realize a collective optical magnetic response and responses to higher-order multipole moments. The deeply subwavelength trapping in the optical lattice allows strong magnetic-dipole and electric-quadrupole interactions.

\begin{figure}[t]
  \centering
   \includegraphics[width=\columnwidth]{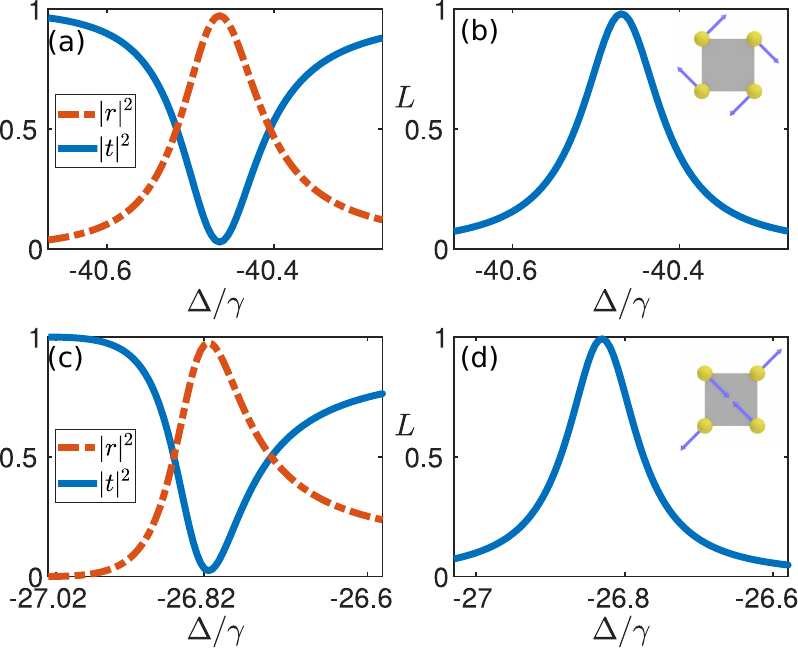}
  \caption{Excitation of the collective magnetic-dipole mode (a,b) and electric-quadrupole mode (c,d) of a bilayer lattice. 
  Electric dipole transitions of the four atoms at the corners of the unit cell approximate a circle around the unit cell, giving rise to a magnetic dipole [(b), inset]. Another configuration, with dipoles pointing towards or away from the center, gives rise to an electric quadrupole [(d), inset]. The collective modes consist of a uniform in-phase oscillation of the magnetic dipole or electric quadrupole on each unit cell. (a) Transmission and reflection of a Gaussian beam as the probe laser detuning is varied through the magnetic-dipole resonance. (b) Occupation of the collective magnetic-dipole mode. (c) Transmission and reflection as detuning is varied through the electric-quadrupole resonance. (d) Occupation of the collective electric-quadrupole mode. The total occupation of all modes in (b,d) is normalized to $1$ on resonance. $21\times 21\times 4$ lattice, incident Gaussian beam waist $w=6\lambda$. }
  \label{fig:mag}
\end{figure}

We focus on a specific implementation of the trapping geometry described in Sec.~\ref{sec:trapping} with $a_x=a_y=0.08\lambda$, $d_z=0.82\lambda$, and $d_y=8a_y$.
Each square unit cell in isolation has a variety of different excitation eigenmodes, each of which is dominated by a different multipole, including electric dipole, magnetic dipole, electric quadrupole, and magnetic quadrupole contributions~\cite{Ballantine20Huygens}. 
Collective interactions between unit cells lead to eigenmodes of the whole array that correspond to uniform repetitions of the unit cell eigenmodes, representing a particular multipole moment. 
Numerically, we find a single unit-cell eigenmode for which the magnetic dipole component on resonance is over 99\%, with negligible contributions from other electromagnetic multipoles (see Appendix~\ref{sec:multipole}). The magnetic response arises as the orientations of the oscillating electric dipoles on each atom form a discrete approximation of a loop [see inset of Fig.~\ref{fig:mag}(b)]. 
A unit cell excitation eigenmode with electric quadrupole dominating radiation is shown in the inset of Fig.~\ref{fig:mag}(d). Here an electric quadrupole constitutes over 88\% of the response, with the rest dominated by an electric dipole (see Appendix~\ref{sec:multipole}).

The transmission and reflection spectra of an incident Gaussian beam close to the resonance of the mode corresponding to uniform in-phase magnetic dipoles on each unit cell are shown in Fig.~\ref{fig:mag}(a). 
The occupation of this mode, defined as~\cite{Facchinetti16}
\begin{equation}
    L_i=\frac{|\vec{v}_i^T\vec{b}|^2}{\sum_j|\vec{v}_j^T\vec{b}|^2},
\end{equation}
is displayed in Fig.~\ref{fig:mag}(b), with the total occupation of all modes normalized at the resonance frequency. The magnetic mode is almost uniquely occupied on resonance, indicating a strong magnetic response. The transmission and reflection spectra of the uniform in-phase electric-quadrupole mode are shown in Fig.~\ref{fig:mag}(c). Figure~\ref{fig:mag}(d) shows the occupation of the corresponding collective mode over the entire lattice, again with almost unit occupancy on resonance.

The deeply subwavelength trapping geometry of bosonic Sr, with $a_x$, $a_y=0.08\lambda$, opens new avenues for engineering and manipulating light-matter interactions since magnetic-dipole and electric-quadrupole moments on each unit cell can themselves form a subwavelength lattice.
For example, the bilayer of bosonic Sr atoms demonstrated here provides an experimental platform to realize previous proposals, such as a magnetic mirror~\cite{Ballantine21wavefront}, which reflects light without the usual $\pi$ phase shift present upon reflection from an electric-dipole array~\cite{Sievenpiper99,Schwanecke06,Liu14}, or unidirectional storage of a single-photon pulse~\cite{Ballantine21bilayer}. In addition, the small lattice spacing scheme of Sr could also be extended to more complex setups involving topologically nontrivial configurations, such as toroidal dipoles and anapoles~\cite{Ballantine20Toroidal}.

\section{Atomic Huygens' surface}
\label{sec:HSshifts}

\subsection{Engineering atomic level shifts}

The higher-order multipole response of tightly spaced bosonic Sr arrays can also be used to engineer an atomic Huygens' surface. Here, we describe a physical protocol using the electronic level and transition parameters of bosonic Sr for realizing optical wavefront control via a Huygens' surface. This is achieved by engineering the Sr atomic level shifts so that both electric-dipole and magnetic-dipole modes, or alternatively both electric-dipole and electric-quadrupole modes, are simultaneously excited.
We calculate the effect of an additional off-resonant optical standing wave to synthesize the required level shift profiles and numerically find generally slightly better solutions for the electric-dipole and electric-quadrupole mode pair.
Our scheme, therefore, provides a precise implementation of a bosonic Sr Huygens surface which, by varying the transmitted phase across the array, could realize previous proposals for wavefront control of incident light, including focusing, beam-steering, polarization control, and optical angular momentum generation~\cite{Ballantine20Huygens,Ballantine21wavefront,Ballantine21quantum}. Huygens' surfaces have been engineered in metallic~\cite{Pfeiffer13} and dielectric~\cite{Decker15,Yu15,Chong15,Shalaev15} systems. However, high absorption and fabrication imperfections pose severe challenges at optical frequencies. All absorbed light is re-radiated for the bosonic Sr atomic array considered here (unit emission quantum yield), and each atom has an identical resonance frequency. The atomic response could also be manipulated at the quantum regime~\cite{Ballantine21quantum}.

While the Huygens' surface effect is typically produced by a combination of electric and magnetic dipoles, we generate it using electric dipoles and electric quadrupoles. 
 The symmetric emission of the electric dipole moment in the forward and backward directions, combined with the antisymmetric emission of the electric-quadrupole moment, leads to destructive interference in the reflected light and constructive interference in transmission. 
We write the scattered field $\vec{E}_{s}^{\pm}$ in the $\pm\unitvec{x}$ directions as $\vec{E}_s^{\pm}=\vec{E}_{s,d}^{\pm}+\vec{E}_{s,q}^{\pm}$, where $\vec{E}_{s,d}^{\pm}$ ($\vec{E}_{s,q}^{\pm}$) is the field from the electric-dipole (electric-quadrupole) mode. Then, on resonance, $\vec{E}_{s,d}^{+}=\vec{E}_{s,q}^{+}\approx-\cbE$ and the total field $\vec{E}=\vec{E}_s^{+}+\cbE\approx-\cbE$. 
Tuning the incident light off-resonance leads to transmission with no phase shift. As the resonance is crossed, complete $2\pi$ phase control of the forward-propagating beam is achieved. Meanwhile, in the backward direction, the different symmetries of the scattered fields from the electric dipole and quadrupole contributions ensure $\vec{E}_{s,d}^{-}\approx-\vec{E}_{s,q}^{-}$, leading to near-zero reflection.

\begin{figure}[ht]
\includegraphics[width=0.5\textwidth]{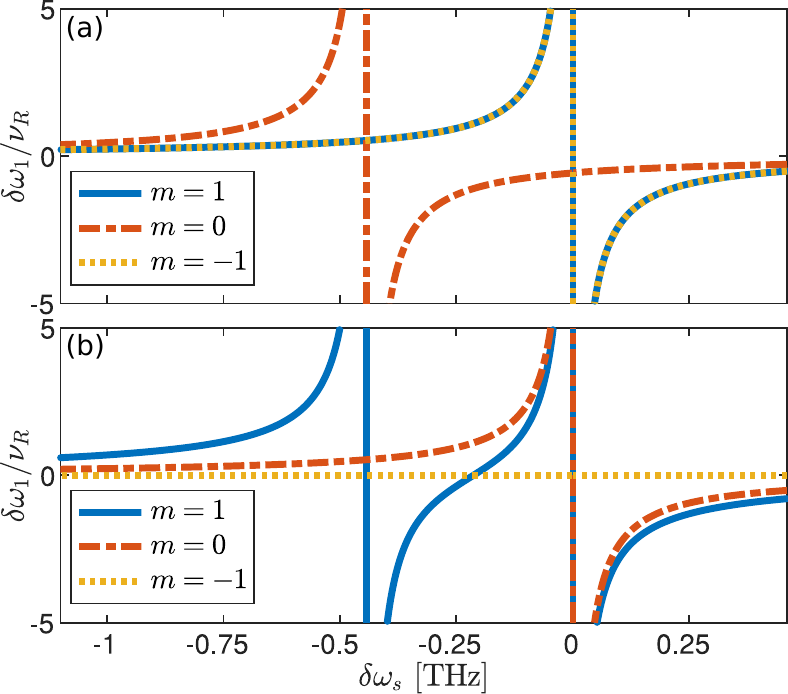}
\caption{Light shift in the units of the recoil energy $\nu_R=h/(2m\lambda_c^2)\approx 5.6\,$kHz, with $\lambda_c=636.39$nm for the three Zeeman substates of the $\,^3D_1$ level of bosonic Sr. (a) $\pi$ polarized light and (b) $\sigma^{+}$ polarized light for $I=1{\rm W}/{\rm cm}^2$ close to resonance with the $\,^3D_1\rightarrow\,^3P_J$ transitions in Table.~\ref{table}. 
} \label{P01} 
\end{figure}

 The collective eigenmodes generally all have different line shifts and linewidths, making their simultaneous excitation with equal scattering strength nontrivial. To excite two modes simultaneously, we vary the shifts of the $m=0,\pm1$ levels between atoms in each unit cell with the same periodicity as the lattice so that different unit cells experience the same level shifts.
Using Eq.~\eqref{eq:LF}, we induce a vector and tensor polarizability (see Appendix~\ref{sec:shifts}) via a control laser close to resonance with the $5s4d\,^3D_1\rightarrow 5s6p\,^3P_{0,1,2}$ transitions described in Table~\ref{table}.
The transition $\,^3D_1\rightarrow\,^3P_2$ is weak and will be ignored, whereas the other far-detuned dipole-allowed transitions will contribute to an extra $m-$independent shift that can be easily measured and compensated for experimentally. The latter will be ignored as well. In Fig.~\ref{P01}(a) [Fig.~\ref{P01}(b)] we show the light shift of the $\,^3D_1$ Zeeman substates across the transition for a plane wave of intensity $1\,\textrm{W/cm}^2$ and a light polarization of $\pi$ ($\sigma^+$). The level shift due to $\sigma^-$ polarization is obtained by reversing the $m=\pm 1$ states.

\begin{table}[ht]
\caption{Wavelengths and linewidths of selected $5s4d\,^3D_1\rightarrow 5s6p\,^3P_{0,1,2}$ transitions used for controlling level shifts.} 
\centering 
\begin{tabular}{c c c} 
\hline\hline 
Transition & Wavelength ($nm$) & $\gamma$ ($s^{-1}$) \\ [0.5ex] 
\hline 
$^3D_1\rightarrow\,^3P_0$ & $636.99$ & $9\times 10^{6}$ \\ 
$^3D_1\rightarrow\,^3P_1$ & $636.39$ & $1.85\times 10^{6}$ \\ 
$^3D_1\rightarrow\,^3P_2$ & $636.12$ & $4\times 10^{4}$ \\ [0.5ex] 
\hline 
\end{tabular}
\label{table}
\end{table}

To engineer the necessary level shifts, we employ a control beam that consists of a standing wave with a wavelength close to the transitions in Table~\ref{table} and tilted at an angle such that the in-plane periodicity matches that of the atomic lattice. 
The control beam has the form $\cbE_c(\vec{r})=\mathcal{E}_c\unitvec{e}_c\cos{(\vec{k}_c\cdot\vec{r}+\phi)}$, with amplitude $\mathcal{E}_c$, frequency $\omega_c = 2\pi c/(636.39{\rm nm})+\delta\omega_c$, and wavevector $\vec{k}_c$.  
The wavevector $\vec{k}_c=(0.64\unitvec{x}+0.77\unitvec{y})k_c$, where $k_c=|\vec{k}_c|=\omega_c/c$, is fixed to give an in-plane periodicity $d_y/2$. This leaves a number of free parameters numerically optimized to maximize the transmission. The control beam has freedom of phase $\phi$, intensity $I_c=c\epsilon_0|\mathcal{E}_c|^2/2$, frequency $\delta\omega$, and the orientation and ellipticity of the polarization $\unitvec{e}_c$. In addition, the lattice parameters $a_x$, $a_y$, and an overall constant Zeeman splitting $\bar{\delta}=(\Delta_{+}^{(j)}-\Delta_{-}^{(j)})/2$ (achievable, e.g., with a constant magnetic field) are varied. This results in eight parameters to control the seven independent relative level shifts $\Delta_{\pm}^{(j)}$ on four atoms, up to an irrelevant overall shift in the frequency of the Huygens' surface resonance, allowing for redundancy to account for experimental errors or constraints. 

\begin{figure}[htbp]
  \centering
   \includegraphics[width=\columnwidth]{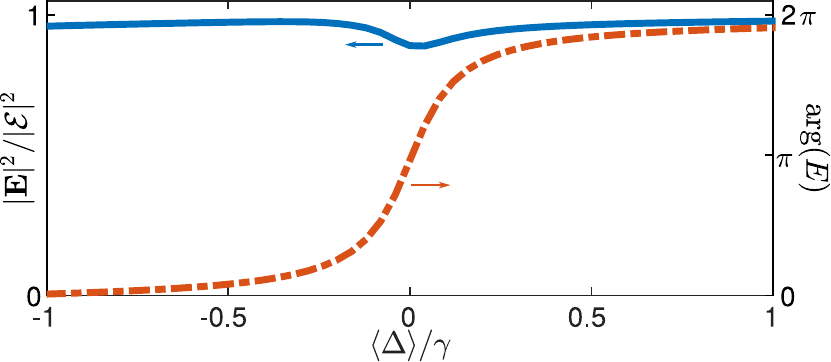}
  \caption{Intensity (left axis) and phase (right axis) of light transmitted through an atomic Huygens' surface engineered using the parameters of bosonic Sr. A complete $2\pi$ phase range is achieved with transmission $\geq 89\%$. Plane-wave illumination of a $31\times 31\times 4$ array of atoms trapped at the magic wavelength $a_x=a_y=0.08\lambda$, with $d_z=0.82\lambda$ and $d_y=8a_y$. The control beam parameters are: $\bar{\delta}=-13\gamma$, $I_c=2.36{\rm kW}/{\rm cm}^2$, $\delta\omega=2.7\times10^{10}$Hz, and $\phi=1.1$. 
   }
  \label{fig:transmission}
\end{figure}

While numerous solutions are possible, here we take $\bar{\delta}=-13\gamma$, $I_c=2.36{\rm kW}/{\rm cm}^2$, and $\delta\omega=2.7\times10^{10}$Hz, slightly blue-detuned from the $^3\,D_1\rightarrow\,^3P_1$ transition where level shifts are large, but still far enough from resonance to disregard spontaneous emission events. The phase $\phi=1.1$ gives a large variation in intensity across a unit cell, while the polarization $\unitvec{e}_c=(0.7 + 0.2i)\unitvec{x}+(-0.6 - 0.2i)\unitvec{y}+0.03\unitvec{z}$ is chosen to provide significant scalar, vector, and tensor polarizabilities (see Appendix~\ref{sec:shifts}). 
The resulting Huygens' surface transmission spectrum is shown in Fig.~\ref{fig:transmission} for a plane-wave incident beam with uniform $y$ polarization. Numerically, we find transmission is $\gtrsim 89\%$ across the full $2\pi$ phase range.

\begin{figure}[htbp]
  \centering
   \includegraphics[width=\columnwidth]{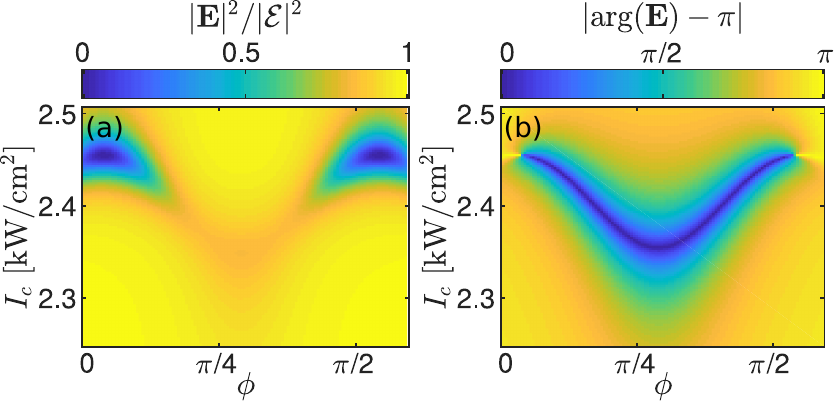}
  \caption{Compensating for experimental variation in Huygens' surface parameters. (a) Transmission of light incident on the array, and (b) deviation from the targeted $\pi$ phase shift for the resonant incident field, as the phase of the control beam is allowed to vary ($x$ axis). 
  High transmission with the desired phase can be maintained by adjusting the control beam intensity ($y$ axis). All other parameters are kept fixed. The transmission for a $\pi$ phase shift corresponds to the minimum transmission in the spectrum (see Fig.~\ref{fig:transmission}). Plane-wave illumination with spacing as in Fig.~\ref{fig:transmission}.
   }
  \label{fig:errors}
\end{figure}

While the solution provided depends sensitively on each individual parameter, multiple similar solutions exist with different parameter values. Therefore, experimental uncertainty in one parameter can be compensated for by reoptimizing the rest. We illustrate this with an example shown in Fig.~\ref{fig:errors}, corresponding to the case analyzed in Fig.~\ref{fig:transmission}. We consider the ``worst-case scenario'' of the laser set to target the resonance where the transmission is the lowest and the gradient of the phase of transmitted light largest (with respect to the laser frequency). At the exact resonance, the Huygens' surface generates a $\pi$ phase shift in the transmitted light, but the variation of the phase $\phi$ of the control beam from $\phi\simeq 1.1$ alters this. Keeping all other parameters fixed, we find a wide range of $\phi$ where an optimal solution can be restored by also varying the intensity $I_c$ of the control beam, achieving the desired phase with $\gtrsim 80\%$ transmission for $0.57\lesssim \phi\lesssim 1.26$. This figure describes the minimum transmission in the spectrum for each solution (see Fig.~\ref{fig:transmission}). We also note, at $I_c\approx 2.35{\rm kW}/{\rm cm}^2$, the solution becomes insensitive to small fluctuations in $\phi$.
Other experimental errors can be compensated for by allowing more parameters to be reoptimized. E.g., a mismatch in the detuning $\delta\omega$, which changes the respective magnitude of the scalar, vector, and tensor polarizabilities, can be compensated for by simultaneously varying both the polarization and the intensity of the control beam.

\subsection{Examples of wavefront control}

The Huygens' surface can be used for arbitrary wavefront shaping by applying a spatial variation to the transmission phase across the surface. 
This variation is chosen to reconstruct the desired propagating wave, with the incident field amplitude $\mathcal{E}$ transformed to the outgoing field $\exp{[i\alpha(y,z)]}\mathcal{E}$. Here, we demonstrate two examples of wavefront shaping; beam steering and preparation of a Poincar\'{e} beam, using the transitions of a bosonic Sr atom Huygens' surface.

\subsubsection{Beam steering}

\begin{figure}[htbp]
  \centering
   \includegraphics[width=\columnwidth]{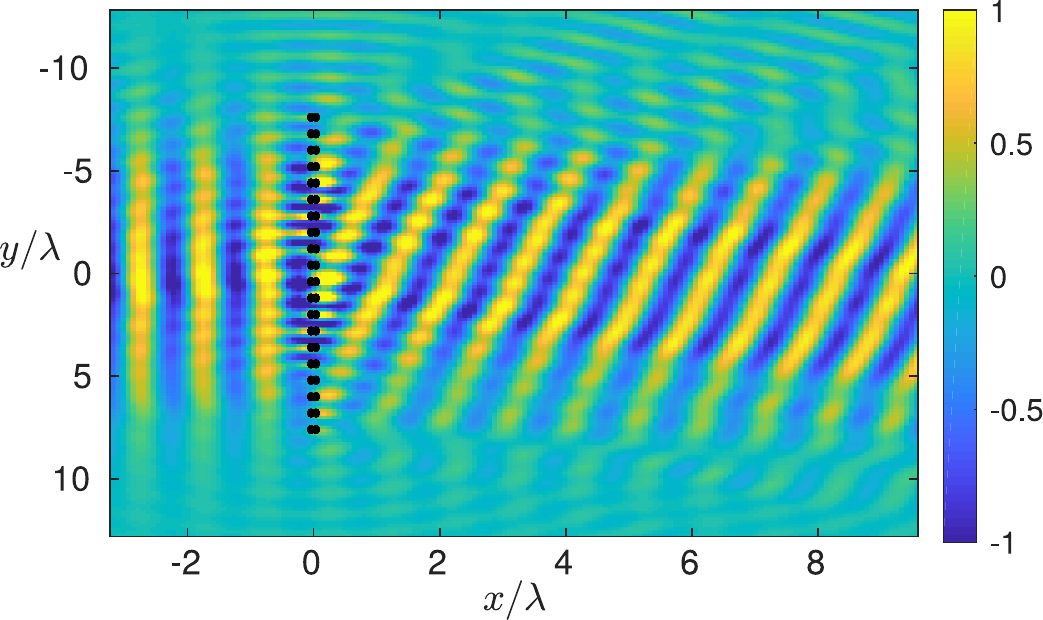}
  \caption{Beam-steering of a Gaussian beam, normally incident on the array, by an angle of $10^\circ$. Real part of the electric field $\mathrm{Re}(E_y)/|\mathcal{E}(0)|$, incident from the left, transmitted through the array at $x=0$ (illustrated by black dots) with a linear phase profile in the $y$ direction, leading to steering in the $xy$ plane. $21\times 21\times 4$ array with spacing as in Fig.~\ref{fig:transmission}, and incident Gaussian beam waist $w=6\lambda$.
  }
  \label{fig:bs}
\end{figure}

A beam of light incident on the atomic Huygens' surface at one angle can be steered under the transmission to leave the array at a different angle~\cite{Liu17bs}. This mechanism provides a vital resource to redirect light, e.g., for subsequent beam shaping or detection. 
The steering is achieved by applying a linear phase gradient, $\alpha=\kappa_y y+\kappa_z z$. For example, a normally incident field with phase $\exp{(i k_x x)}$ is transformed to the outgoing field with phase $\exp{[i(k_x^\prime x+\kappa_y  y+\kappa_z z)]}$, with effective momentum components $k_y=\kappa_y$, $k_z=\kappa_z$, and $(k_x^\prime)^2=k_x^2-(\kappa_y^2+\kappa_z^2)$. The resulting beam is steered away from the x axis by an angle $\sin^{-1}(\kappa/k)$, where $\kappa=\sqrt{\kappa_y^2+\kappa_z^2}$. Figure~\ref{fig:bs} shows an example of steering in the $y$ direction, $\kappa_z=0$, by an angle $\theta=10^\circ$. The desired phase profile is given by
\begin{equation}
\kappa_y = \frac{2\pi}{\lambda}\sin{\theta}.    
\end{equation}
Because the transmission phase shift in Fig.~\ref{fig:transmission} depends on the detuning of the laser from the Huygens' surface resonance and not the relative shifts between the $m=0,\pm 1$ levels, the wavefront shaping can be achieved by adding a simple uniform phase gradient. For example, scalar polarizability generated by an additional off-resonant optical or microwave field propagating in the $x$ direction with polarization $\sqrt{2/3}\unitvec{e}_y+\sqrt{1/3}\unitvec{e}_z$ and with an intensity gradient in the transverse plane is sufficient. For the example in Fig.~\ref{fig:bs}, the required level shift gradient in the $y$ direction is $0.17\gamma/\lambda=0.07\gamma\, \mu{\rm m}^{-1}$.
In contrast to fabricated plasmonic or dielectric devices, beam steering angle could thus be changed in situ by varying the gradient or direction of the intensity variation.

\subsubsection{Poincar\'{e} beam}

We now show how the atomic Huygens' surface can be used to transform a Gaussian incident beam with uniform polarization into a Poincar\'e beam, for which the Stokes vector of the field amplitude spans the entire surface of the Poincar\'{e} sphere in a single transverse cross-section~\cite{Beckley10}. The Poincar\'{e} beam amplitude is formed from a linear combination,
\begin{equation}
\label{eq:poincare}
\textbf{U}_P = U_{00}\unitvec{e}_z+U_{10}\unitvec{e}_y,    
\end{equation}
of a Gaussian and Laguerre-Gauss (LG) beam with orthogonal polarizations. Here, $U_{lp}$ is the LG beam with integer azimuthal index $l$ and radial index $p$, and an angular phase profile $e^{il\phi}$ where $\phi=\arctan{(k_z/k_y)}$~\cite{Forbes21}. The different radial amplitude profile and azimuthal phase profile of the two components interfere, giving varying polarization of $\vec{U}_P$ across the transverse plane of the beam.

To achieve this superposition with an atomic Huygens' surface, we illuminate it with an incident Gaussian with polarization $\unitvec{e}_{\rm in}=0.8\unitvec{e}_y+0.6\unitvec{e}_z$. As the unit cell electric-dipole and electric-quadrupole modes consist of atomic dipoles in the $xy$ plane, the $z$ component does not excite the Huygens' surface and is transmitted with no phase change (other $z$ polarized modes exist but are far off-resonance). 
A phase profile $\alpha=\phi$ on the $y$ component leads to a zero along the $x$ axis and an approximate Laguerre-Gauss profile with $l=1$, with the resulting combination having the form of Eq.~(\ref{eq:poincare}).
The required scalar level shifts can once again be induced with an additional optical field, now with an azimuthal intensity variation. For example, passing light through a surface with an optical thickness proportional to angle, such as a variable neutral density filter, with a discontinuity at a specific angle leads to a corresponding azimuthal phase profile in the transmitted $y$ polarized light.

\begin{figure}[htbp]
  \centering
   \includegraphics[width=\columnwidth]{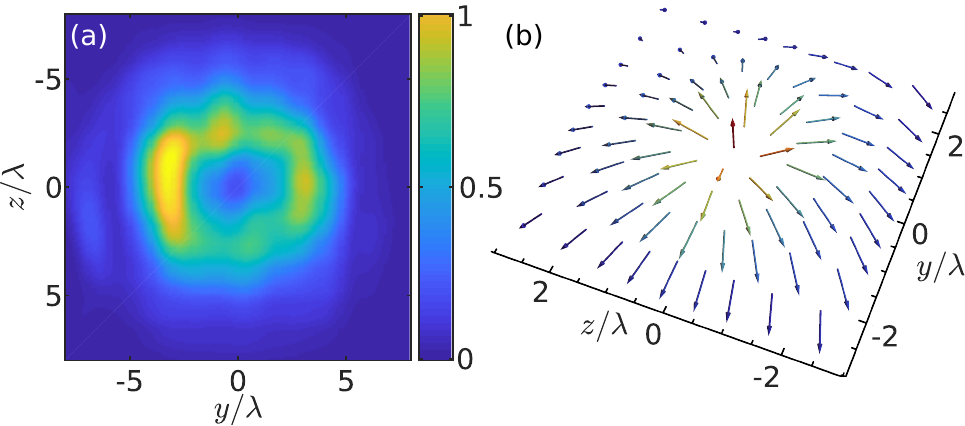}
  \caption{Generation of a Poincar\'{e} beam containing all possible polarizations in a single cross-section. (a) Intensity $I$ of the Poincar\'{e} beam in the $yz$ plane at $x=15\lambda$. (b) The Stokes vector $\vec{S}$ showing baby-Skyrmion structure, with the characteristic fountain-like profile. At the center of the beam, the light is $z$ polarized ($s_3\approx 1$). Some distance from the center, the magnitude of the $y$ and $z$ polarization components are equal. Here, $\vec{S}$ lies in the plane and points radially, corresponding to diagonal, left circular, anti-diagonal, and right circular polarization as the relative phase of the polarization components varies. Far from the center, the light is $y$ polarized ($s_3\approx -1$). 
  $21\times 21\times 4$ lattice with incident Gaussian beam waist $w=6\lambda$ and spacing as in Fig.~\ref{fig:transmission}.
  }
  \label{fig:poincare}
\end{figure}

The intensity and Stokes parameters of the resulting Poincar\'{e} beam are shown in Fig.~\ref{fig:poincare}. For optical fields, the $S^2$ Poincar\'e sphere is parameterized by the Stokes vector ${\bf S}$~\cite{BOR99}. Here, we plot $\vec{S}=(s_1,s_2,s_3)^T$ with a non-standard basis $s_1=(U-V)/\sqrt{2}$, $s_2=(U+V)/\sqrt{2}$, and $s_3=-Q$, where $Q=(|E_y|^2-|E_z|^2)/|E|^2$, $U=2{\rm Re}(E_yE_z^\ast)/|E|^2$, and $V=-2{\rm Im}(E_yE_z^\ast)/|E|^2$ are the standard normalized Stokes parameters describing horizontal, diagonal, and circular polarisation, respectively.
For the $y$ polarized $U_{10}$, the amplitude is zero along the beam axis where the phase is undefined, and the remaining field consists solely of the $z$ polarized $l=0$ component. Moving out from the beam axis, the amplitude of $U_{10}$ increases, and, at some distance, is equal to that of $U_{00}$. The $e^{i\phi}$ phase of the $y$ polarized light then leads to diagonal ($\phi=0$), left-circular ($\phi=\pi/2$), anti-diagonal ($\phi=\pi$), and right-circular ($\phi=3\pi/2$) polarizations around this contour. The $l=0$ component has a narrower spatial profile in the $yz$ plane than the $l=1$ component, and far from the optical axis, the beam is $y$ polarized.  

The preparation of the Poincar\'e beam also represents a topological transformation of a (topologically) trivial beam to one that forms a baby-Skyrmion texture~\cite{Donati2016,Gao2020}. For a baby-Skyrmion, the Stokes vector is assumed to take a uniform constant value everywhere sufficiently far away from the center of the texture. The baby-Skyrmion topology is then determined by a $S^2\rightarrow S^2$ mapping, indicating how many times the $S^2$ compactified beam cross-section wraps over the $S^2$ Poincar\'e sphere,
\begin{equation}\label{Eq:SkyrmionNumber}
W =\int_{\mathcal{S}}\frac{d\Omega_i}{8\pi} \epsilon_{ijk} \textbf{S}\cdot\frac{\partial\textbf{S}}{\partial r_j}\times \frac{\partial \textbf{S}}{\partial r_k},
\end{equation}
where the integration is over the beam cross-section containing the texture and $\epsilon_{ijk}$ denotes a completely antisymmetric Levi-Civita tensor. The field configuration in Fig.~\ref{fig:poincare} corresponds to a nontrivial topological winding number $W=1$, in which case the Stokes vector profile of the beam cross-section covers the Poincar\'e sphere exactly once. Baby-Skyrmions, different from the full 3D Skyrmions~\cite{Skyrme1961,manton-sutcliffe,Sugic21}, can optically appear as a result of different light-matter interfaces~\cite{Tsesses2018, Du2019, Davis2020,Parmee22} and are known in magnetic structures~\cite{Nagaosa2013} and superfluids~\cite{ leanhardt_prl_2003,choi_prl_2012,weiss_ncomm_2019}.

\section{Concluding Remarks}

Bosonic Sr optical lattices provide an exciting platform for studying cooperative interactions of atoms with light, featuring a $J=0\rightarrow J^\prime=1$ transition and with a magic wavelength allowing for deeply subwavelength trap spacing. Here we have shown how a bilayer optical lattice of bosonic Sr atoms can be realized to synthesize a strong collective magnetic response at optical frequencies and how additional level shifts can be introduced to engineer an atomic Huygens' surface allowing for arbitrary wavelength control. The calculations are performed using the parameters of electronic levels and transitions of bosonic Sr.

While we presented a detailed proposal for the trapping and control beam geometries, numerous other solutions using the same experimental platform are also possible, e.g., with larger $a_x$ and $a_y$. Additional improvements in the performance of the Huygens' surface could be achieved by choosing the trapping laser frequency and polarizations, such that the optical lattice lasers also impose level shifts, increasing the control degrees of freedom at the expense of increasing the sensitivity to laser intensity fluctuations.
Similar trapping geometries and site-specific level shifts could also be used to implement many other proposed schemes, including an atomic magnetic mirror~\cite{Ballantine21wavefront}, unidirectional storage of a photon pulse~\cite{Ballantine21bilayer}, or an excitation of a highly subradiant antiferromagnetic mode~\cite{Ballantine20ant}.

Data used in the publication is available at (DOI TO BE ADDED IN PROOF).

\begin{acknowledgments}
K. E. B. and J. R. acknowledge financial support from the United Kingdom Engineering and Physical Sciences Research Council (EPSRC) (Grants No. EP/S002952/1, No. EP/P026133/1, and No. EP/W005638/1). D. W. acknowledges the Singapore Ministry of Education and the Centre for Quantum Technologies (CQT/MoE Grant No. R-710-002-016-271).
\end{acknowledgments}

\appendix
\section{Rotational decomposition of level shifts}
\label{sec:shifts}

Equation~(\ref{eq:LF}) gives the general form for the level shift of a transition for light propagating along the quantization axis. We also employ a variety of trapping and control lasers in various other directions. The level shift for a field $\cbE_c \cos{(\vec{k}_c\cdot+\vec{r}+\phi)}\unitvec{e}$ with general polarization $\unitvec{e}$ and direction $\vec{k}_c$ at position $\vec{r}_j$ on level $\mu$ can be calculated from the geometric decomposition~\cite{Schmidt16,Rosenbusch09,LeKien13}, 
\begin{align}
\delta\omega_{\mu }^{(j)} &= -\frac{1}{4}|\mathcal{E}_c|^2\cos^2{(\vec{k}_c\cdot\vec{r}_j+\phi)} \left[\vphantom{\frac{(3\mu^2-1)}{2}}\alpha^s(\omega)\right. \nonumber \\ &\phantom{==}+ \left. C \frac{\mu}{2} \alpha^v(\omega)    -D\frac{(3\mu^2-2)}{2}\alpha^T(\omega)\right], \label{eq:shifts}
\end{align} 
where $C=|e_{-}|^2-|e_{+}|^2$ parameterizes the degree of circular polarization, $D=1-3|e_0|^2$, and $\alpha^{\rm s}(\omega)$, $\alpha^{\rm v}(\omega)$, and $\alpha^{\rm T}(\omega)$ are the scalar, vector, and tensor polarizabilities, respectively.
The rotational decomposition of the polarizabilites of the $J^\prime=1$ levels can be related to the direct calculation of the level shift $\delta\omega_{m,q}$ of level $m$ with light polarization $q$ by
\begin{align}
\alpha^s &\propto (\delta\omega_{1,0}+\delta\omega_{1,-1})/3,\\
\alpha^v &\propto \delta\omega_{1,-1},\\
\alpha^T &\propto (2\delta\omega_{1,0}-\delta\omega_{1,-1})/3.
\end{align}

\section{Trapping geometry}
\label{sec:trap}

The trapping geometry consists of three independent standing waves in the $x$, $y$, and $z$ directions.  
The standing wave in the $x$ direction is given by  
$\mathcal{E}_m\cos{(2\pi x/\lambda_m)}\unitvec{e}_y$
with a similar standing wave in the $y$ direction obtained by interchanging $x,y\rightarrow y,z$
The trapping in the $z$ direction is achieved with two tilted beams,
$\mathcal{E}_m\left[\unitvec{e}_{1}e^{i(k_y y+k_z z)}+\unitvec{e}_{2}e^{i(k_y y-k_z z)}\right]$, with $k_z=0.6k_m$ and $k_y=\sqrt{k_m^2-k_z^2}$,
and polarizations $\unitvec{e}_{1}=(-k_z\unitvec{e}_y+k_y\unitvec{e}_z)/k_m$ and $\unitvec{e}_{2}=(k_z\unitvec{e}_y+k_y\unitvec{e}_z)/k_m$.
Interference between the beams can lead to a vector or tensor polarizability, leaving the shifts sensitive to fluctuations in the trapping intensity. To avoid this, each standing wave is detuned by $100$MHz, and the average polarization in each direction then leads to a level independent shift on each of the $m=0,\pm 1$ levels of the $^3D_1$ state. 

\section{Multipole expansion of far-field scattering}
\label{sec:multipole}

In Sec.~\ref{sec:magnetism}, we describe uniform collective magnetic-dipole and electric-quadrupole modes and give the contribution of the multipole moments to the corresponding single unit-cell eigenmode. The scattered field at large distances from a subwavelength source can be expanded as
\begin{equation}
\vec{E}_s = \sum_{l=0}^\infty\sum_{m=-l}^{l}\left( \alpha_{E,lm} \vecg{\Psi}_{lm}+\alpha_{B,lm}\vecg{\Phi}_{lm}\right),
\end{equation}
where $\vecg{\Psi}_{lm}$, $\vecg{\Phi}_{lm}$ are the vector spherical harmonics and $\alpha_{E,lm}$ ($\alpha_{B,lm}$) are the electric (magnetic) dipole, quadrupole, etc., components for $l=1,2,\ldots$, respectively~\cite{Jackson}.
For the magnetic-dipole mode described in Fig.~\ref{fig:mag}(a,b), the single unit-cell eigenmode shown in the inset of Fig.~\ref{fig:mag}(b) has $\sum_m|\alpha_{B,1m}|^2\agt 0.99$. For the electric-quadrupole mode described in Fig.~\ref{fig:mag}(c,d), meanwhile, the dominant contribution to the single unit-cell eigenmode shown in the inset of Fig.~\ref{fig:mag}(b) is $\sum_m|\alpha_{E,2m}|^2\agt 0.88$, with an electric-dipole contribution accounting for almost all of the remainder.

\end{document}